\documentclass[a4paper,fleqn]{cas-dc}

\usepackage[numbers]{natbib}
\usepackage{graphicx}
\usepackage{amsmath}
\usepackage{array}
\usepackage{caption}  
\usepackage{booktabs}
\usepackage{tabularx}
\usepackage{multirow}
\usepackage{hyperref}
\usepackage{placeins}
\usepackage{orcidlink}
\usepackage{float}
\usepackage{gensymb}
\usepackage{enumitem}
\usepackage{hyperref}
\usepackage{breakurl}
\usepackage[intoc]{nomencl}
\usepackage{tcolorbox}
\usepackage{xurl}
\makenomenclature
\usepackage{siunitx}
\usepackage{makecell}
\usepackage{placeins}

\newcolumntype{L}[1]{>{\raggedright\arraybackslash}p{#1}}
\newcolumntype{C}[1]{>{\centering\arraybackslash}p{#1}}
\newcolumntype{Y}{>{\centering\arraybackslash}X}

\begin{document}

\shorttitle{Socio-Spatial Contagion Dynamics in Off-Grid PV Adoption}

\shortauthors{Blushtein-Livnon et~al.}

\title [mode = title]{From Expansion to Consolidation: Socio-Spatial Contagion Dynamics in Off-Grid PV Adoption}

\author[1]{Roni~Blushtein-Livnon\orcidlink{0000-0002-3493-4894}}
\fnmark[1]
\author[1,2]{Tal~Svoray\orcidlink{0000-0003-2243-8532}}
\author[3]{Itay~Fischhendler\orcidlink{0000-0001-5355-9316}}
\author[4]{Havatzelet~Yahel\orcidlink{0000-0002-9956-0491}}
\author[4]{Emir~Galilee\orcidlink{0000-0001-7892-4172}}

\fntext[1]{Corresponding author.
E-mail address: livnon@bgu.ac.il}

\affiliation[1]{organization={Department of Environmental, Geoinformatics and Urban Planning Sciences, Ben-Gurion University of the Negev}, country={Israel}}
\affiliation[2]{organization={Department of Psychology, Ben-Gurion University of the Negev}, country={Israel}}
\affiliation[3]{organization={Department of Geography, The Hebrew University of Jerusalem}, country={Israel}}
\affiliation[4]{organization={Ben Gurion Institute for the Study of Israel and Zionism, Ben-Gurion Israel Research Institute}, country={Israel}}

\begin{abstract}
In traditional rural societies, where social ties are tightly embedded in physical space, the diffusion of emerging technologies may be amplified through socio-spatial contagion (SSC). Such processes may play a key role in accelerating residential PV adoption in off-grid regions. Yet empirical evidence on SSC in PV adoption remains largely confined to affluent, grid-connected contexts, limiting its transferability to off-grid settings, where empirical assessment is further constrained by the absence of systematic installation records.
To address these gaps, we leverage a deep learning segmentation model to extract PV installations from a decade-long series of remote sensing imagery across 507 off-grid settlement clusters (hereafter, communities), generating empirical evidence of adoption in data-scarce contexts and enabling data-driven spatio-temporal point pattern inference of SSC. SSC is estimated by quantifying the range and intensity of clustering of new installations around prior adopters. The dynamics of these SSC dimensions were tracked throughout the diffusion process and associated with adoption outcomes. We found that SSC is nearly ubiquitous, often spanning most of the community’s spatial extent, while exhibiting substantial heterogeneity in intensity. Although SSC amplifies over time, its effects remain temporally concentrated, peaking within 1–2 years of nearby installations and attenuating thereafter. Linking SSC intensity to adoption rates reveals a positive association in both cross-sectional and temporal analyses. However, the relationship between SSC range and adoption evolves over time: in early diffusion phases, adoption growth is associated with range expansion, whereas in later phases it aligns with range contraction. This shift reflects a reconfiguration of spatial installation patterns from wide-ranging clustering to consolidation. The role of SSC in shaping PV diffusion underscores the potential of seeding interventions to accelerate adoption, particularly when aligned with contagion dynamics throughout the adoption process.
\end{abstract}

\begin{keywords}
PV Adoption \sep Socio-spatial contagion \sep peer effects \sep off-grid \sep dynamics analysis \sep point-process inference 
\end{keywords}

\maketitle
\section{Introduction}
Understanding the diffusion of solar energy technologies is central to decarbonization efforts, particularly as current transition rates remain insufficient to meet climate stabilization pathways \citep{IRENA_2025}. Residential PV adoption, however, is not uniform, exhibiting pronounced spatial heterogeneity across geographic scales \citep{IEA2024Renewables}. Aggregate-scale approaches risk obscuring place-specific adoption dynamics, underscoring the importance of granular spatial analyses to reveal fine-scale patterns and inform context-sensitive interventions that accelerate energy transition \citep{hartmann2008}.

At fine spatial scales, residential PV adoption reflects household-level decisions shaped by structural constraints \citep{ibegbulam2023}, economic considerations \citep{jacksohn2019drivers}, and social interactions embedded in local spatial contexts \citep{zhang2023neighbourhood}. These interactions manifest most directly through peer influence, which can amplify or attenuate spatial heterogeneity in technology adoption. Such peer effects, commonly conceptualized as socio-spatial contagion (denoted here as SSC), refer to spatially embedded processes through which exposure to prior adopters shapes adoption decisions, with geographic proximity acting as a key organizing dimension  \citep{lengyel2020role, wolske2020peer}.

Residential PV systems constitute a particularly suitable context for examining SSC, as they are durable, highly visible, and spatially fixed, thereby generating strong and persistent exposure \citep{Graziano2015, BaltaOzkan2021}. A growing body of research has quantified SSC in PV adoption; however, this evidence is drawn almost exclusively from affluent, developed, and grid-connected societies. Off-grid communities across much of the Global South, which in this study are predominantly rural and traditional societies, adopt PV technology under fundamentally different conditions than those studied in the existing literature, calling into question whether current findings can be generalized or transferred to these contexts \citep{oliva2025decoding}. Moreover, PV adoption in off-grid settings exhibits substantial heterogeneity in both the intensity and the functional scope of technology use, further underscoring the need for context-sensitive analyses of SSC.

This oversight is consequential given the central role of solar technologies in addressing energy poverty and expanding access to electricity in underserved regions \citep{ESMAP2024, SDG72025}. In contexts characterized by limited formal infrastructure and institutional support, understanding the spatio-temporal dynamics of adoption through locally embedded social pathways is not only a theoretical and empirical question, but also a practical condition for improving the targeting, timing, and effectiveness of decentralized electrification efforts \citep{opiyo2019impacts, kizilcec2021}.

In off-grid societies, where social networks are tightly embedded in physical space, SSC is likely to be particularly pronounced \citep{Putra2018Biogas, Wen2021Clean}. Comparable contagion mechanisms were documented across other technology domains, including agriculture and household clean energy \citep{Matuschke2009, Gu2022Peer}. 

However, whether similar patterns characterize residential PV adoption remains an open question, as the distinctive characteristics of PV systems give rise to both mechanisms that may promote SSC, such as social recognition and observational learning, and constraints that may limit it, including financial and technical barriers. This interplay suggests that SSC in off-grid PV adoption cannot be inferred from other domains and requires explicit empirical investigation.

These considerations expose an important research gap. While SSC in grid-connected PV adoption was systematically quantified across diverse settings, evidence from off-grid contexts remains scarce and largely qualitative (e.g., \citep{TsoeuNtokoane2025, oliva2025decoding}). This constraint is primarily driven by the limited availability of fine-scale adoption data. Consequently, existing research offers only partial insight into the spatial configuration and dynamics of SSC in off-grid settings. Without systematic evidence on SSC patterns and dynamics in these contexts, adoption policies risk being poorly targeted, overlooking the spatial mechanisms that could most efficiently drive diffusion where it is needed most.

This data limitation creates a methodological bottleneck. Recent advances in remote sensing and computer vision offer a promising pathway to overcome these limitations. Deep learning models applied to high-resolution aerial imagery enable the automated curation of small-scale PV installations, including the extraction of precise locations and surface areas \citep{li2025joint, blushtein2026}. When applied to image archives, such approaches facilitate the reconstruction of empirically grounded PV adoption time series at fine spatial resolution, thereby allowing direct quantification of SSC dynamics.

We address these gaps through a methodological framework that combines deep-learning-based extraction of household PV installations from remote sensing imagery with spatio-temporal point-pattern analysis. This approach, which remains largely unexplored in the context of off-grid PV adoption, enables the reconstruction of behavior-based adoption records in contexts where conventional installation data are unavailable. Specifically, we focus on Bedouin settlement-clusters in the Negev Desert in Israel as a case study of off-grid communities. The analysis traces the diffusion of residential PVs over a decade, from the onset of PV penetration in the area, which followed declining PV costs around 2012, through 2022, capturing both early-phase dynamics and their progression into subsequent phases of the adoption process. By doing so, the study provides insight into the challenge of accelerating off-grid solar diffusion in energy-poor settings, where effective scaling depends not only on the availability of technology but also on how adoption spreads within communities.

Accordingly, this paper is structured around three operative objectives:
\begin{enumerate}  [label=\arabic*., leftmargin=*, labelsep=0.5em, itemsep=-0.52pt, topsep=2pt]
  \item To examine the existence of SSC in PV adoption within off-grid communities, and to assess its spatial reach and temporal dynamics. 
  \item To identify distinct SSC patterns and analyze their association with adoption outcomes.
  \item To analyze transitions in SSC patterns over time and their association with adoption dynamics.
\end{enumerate}

\section{Background}
Diffusion theory has long conceptualized the spread of innovation as a contagion-like process, operating through communication and imitation, mediated by mechanisms such as information exchange, observational learning, and norm formation \citep{Rogers2014, bramoulle2020}. Through these mechanisms, exposure to prior adopters can reduce uncertainty and shape perceptions of a technology’s suitability and benefits. At the aggregate level, these micro-level processes give rise to system-level diffusion dynamics, a perspective reinforced by subsequent work in socio-technical systems that frames societies as learning systems and formalizes diffusion through infection-based models \citep{Marchetti1980, Vespignani2012}. Both interaction pathways and observational exposure are embedded in physical space, rendering contagion processes inherently spatial and positioning geographic proximity as a central organizing dimension of diffusion \citep{lengyel2020role}. 

Building on this perspective, empirical studies of residential PV adoption have focused on the spatial configuration of prior installations, consistently documenting that SSC in PV adoption is substantial and highly sensitive to distance, time, social context, and local setting. Using data from Seattle and Vermont, \citet{Min2025} show that PV exhibits the broadest spatial reach of peer effects among renewable energy technologies, extending beyond 700 m in urban areas and up to 2 km in less densely populated settings. In suburban Connecticut, \citet{Bollinger2022} have found that visible PV installations influence adoption decisions within 500 m, increasing a household’s monthly adoption probability by 9.8\%, an effect comparable to a price reduction of $\sim$\$577. Focusing on urban neighborhoods in the northeastern US, \citet{Graziano2019} demonstrate that spatial spillovers are strongest within ${\sim}0.5$ miles and over short temporal windows of up to six months, decaying rapidly with distance and time. Evidence from less mature markets with mixed urban-rural settings in Poland suggests smaller, but still significant effects; \citet{Sokolowski2023} have found that a single additional PV installation increases subsequent adoption probability by 0.06 percentage points in the following month, and suggested that peer effects can emerge very early in the diffusion process, without requiring a critical mass of installations. Beyond spatial extent and magnitude, SSC was found to be socially heterogeneous. By analyzing census-tract data across diverse US settlement types, \citet{OShaughnes2023} report substantially stronger peer responses among higher-income households than lower-income households (1.6 versus 0.2 percentage points), with within-group peer effects roughly twice as large as cross-group effects. Overall, existing literature provides strong empirical evidence that SSC plays an important role in PV diffusion, while also revealing substantial heterogeneity in its effects and features within grid-connected settings. 

In traditional societies, social interactions are inherently local, rendering SSC a central mechanism for the adoption of new technologies. These social networks are tightly embedded in space, with physical proximity strongly overlapping with kinship ties, repeated everyday interaction, and informal information exchange \citep{Putra2018Biogas, Wen2021Clean}. Unlike grid-connected societies, where PV adoption is decoupled from electricity access and mediated primarily by formal information channels, market institutions, and policy incentives \citep{arnold2022prices, jacksohn2019drivers}, off-grid contexts are far less structured by these mechanisms, making local observation and peer experience the dominant drivers of adoption decisions \cite{Liu2026}. Consequently, the immediate vicinity becomes a channel for the transmission of knowledge, norms, and behavioral cues \citep{Matuschke2009, Zhu2022Peer}, thereby amplifying SSC's role as a central driver of new technology adoption \citep{Conley2010, Gu2022Peer, BandieraRasul2006}.

Nevertheless, empirical studies of SSC in off-grid settings remain scarce for residential PV, even though comparable effects were documented in such contexts for other household-level technologies. Evidence from the adoption of agricultural technologies consistently shows strong spatially proximate SSC. Using detailed network data from rural Mozambique on a newly introduced agricultural crop, \citep{BandieraRasul2006} establish that farmers’ initial adoption decisions are strongly shaped by the adoption behavior of close family and friends, with network effects often larger than socio-economic characteristics such as literacy, household resources, and prior exposure to development interventions. Likewise, \citep{Matuschke2009} found that adoption of hybrid seeds in rural India is primarily driven by exposure to immediate neighbors or tightly bounded village segments rather than village-wide diffusion, with aggregate village measures substantially understating the strength of local peer effects. Recent evidence from China further shows that peer effects weaken rapidly with increasing geographic and economic distance \cite{Xu2022Mutual}, reinforcing the highly localized nature of effective social learning in agricultural technology adoption. 

Consistent with evidence from agricultural technology adoption, research on household clean energy adoption identifies localized social influence, albeit operationalized at a coarser spatial scale. Studies of cleaner cooking and heating technologies typically document the village as the relevant unit of exposure, showing that adoption responds to local prevalence of use and is mediated through repeated observation and conformity to local norms, even after accounting for income, infrastructure, and policy incentives \citep{Zhu2022Peer, Gu2022Peer, Wen2021Clean}.

Yet, it remains unclear whether SSC operates in residential PV adoption in a manner comparable to other technologies, particularly in terms of its spatial scale and strength, given the distinctive characteristics and social properties of PV systems. Two countervailing considerations emerge. One strand of reasoning suggests that SSC in residential PV may be more intense than that observed for other household energy technologies. In off-grid settings, PV systems often function as socially salient assets associated with household self-sufficiency, status, and community recognition \citep{opiyo2019impacts}, thereby reinforcing imitation and social signaling \citep{simpson2021adoption}. Moreover, their high and continuous visibility in compact, often kinship-based settlements, together with frequent informal, face-to-face interactions, may further amplify normative pressure and observational learning \citep{Putra2018Biogas, Wen2021Clean}. As a consequence of this heightened intensity, the effective scale of exposure may extend beyond immediate neighbors. In these communities, often characterized by dense social ties and compact spatial form, exposure may operate at the level of the village as a whole. These features suggest that SSC in PV adoption may be both more intense and more spatially extensive than in other technology domains.

At the same time, adoption of residential PV systems entails substantial upfront investment and technical complexity, which are especially binding in off-grid contexts due to persistent financial scarcity and limited service infrastructure \citep{haliru2022expanding, tamir2015issues}. These constraints may restrict households’ capacity to translate exposure into adoption, even when informational barriers are reduced, thereby attenuating SSC. The interaction between reinforcing and inhibiting forces may therefore result in heterogeneous and fragmented PV diffusion dynamics.

This tension aligns with broader evidence from the PV adoption literature, which documents pronounced heterogeneity in SSC. Meta-analytic findings show that estimated peer effects in PV adoption vary widely across institutional settings and spatial scales \citep{best2023meta}, underscoring that SSC cannot be treated as a universal or monotonic process. These considerations suggest that the strength, spatial reach, and dynamics of SSC in off-grid PV adoption are theoretically ambiguous and therefore require explicit empirical investigation rather than inference from grid-connected contexts or other household technologies.

Only a limited number of studies have begun to examine PV SSC processes in rural traditional contexts. For instance, \citet{Liu2026} demonstrate that spatial peer effects intensify when geographic proximity overlaps with kinship-based ties in rural China. In this context, PV adoption becomes socially observable and embedded within local systems of trust, recognition, and normative expectations, albeit within grid-connected settings. Complementary evidence from off-grid regions in Zambia suggests that adoption dynamics are also shaped by socially embedded learning processes, in which households observe and respond to neighbors' and kin's adoption practices even under severe financing constraints and limited access to formal credit or subsidy schemes \citep{Chanda2025}.

Nonetheless, research in off-grid settings remains constrained in both empirical scope and methodological rigor. Much of the available evidence is based on interviews \citep{WallBake2025, TsoeuNtokoane2025} or cross-sectional surveys \citep{Tetteh2022, Aarakitl2021}, and frequently relies on self-reported intentions rather than observed adoption behavior \citep{best2023meta, Mahn2024, oliva2025decoding}. These limitations largely reflect the scarcity of fine-scale spatio-temporal records of PV installations in these regions \citep{best2023meta, Mahn2024, opiyo2019impacts}. In the absence of such data, rigorous examination of SSC remains challenging, as temporal sequencing and spatial clustering are essential for disentangling peer influence from broader diffusion dynamics \citep{bramoulle2020, graham2018, wolske2020peer, bollinger2012}. Hence, the literature rarely captures how SSC varies in timing, intensity, and spatial reach across different phases of the diffusion process, offering only partial insight into its underlying dynamics. We address these limitations here by providing a fine-scale, spatiotemporal analysis of SSC dynamics based on observed off-grid PV adoption behavior.

\section{Methodology}
\begin{figure*}[H!]
    \centering
    \includegraphics[width=1
    \textwidth]{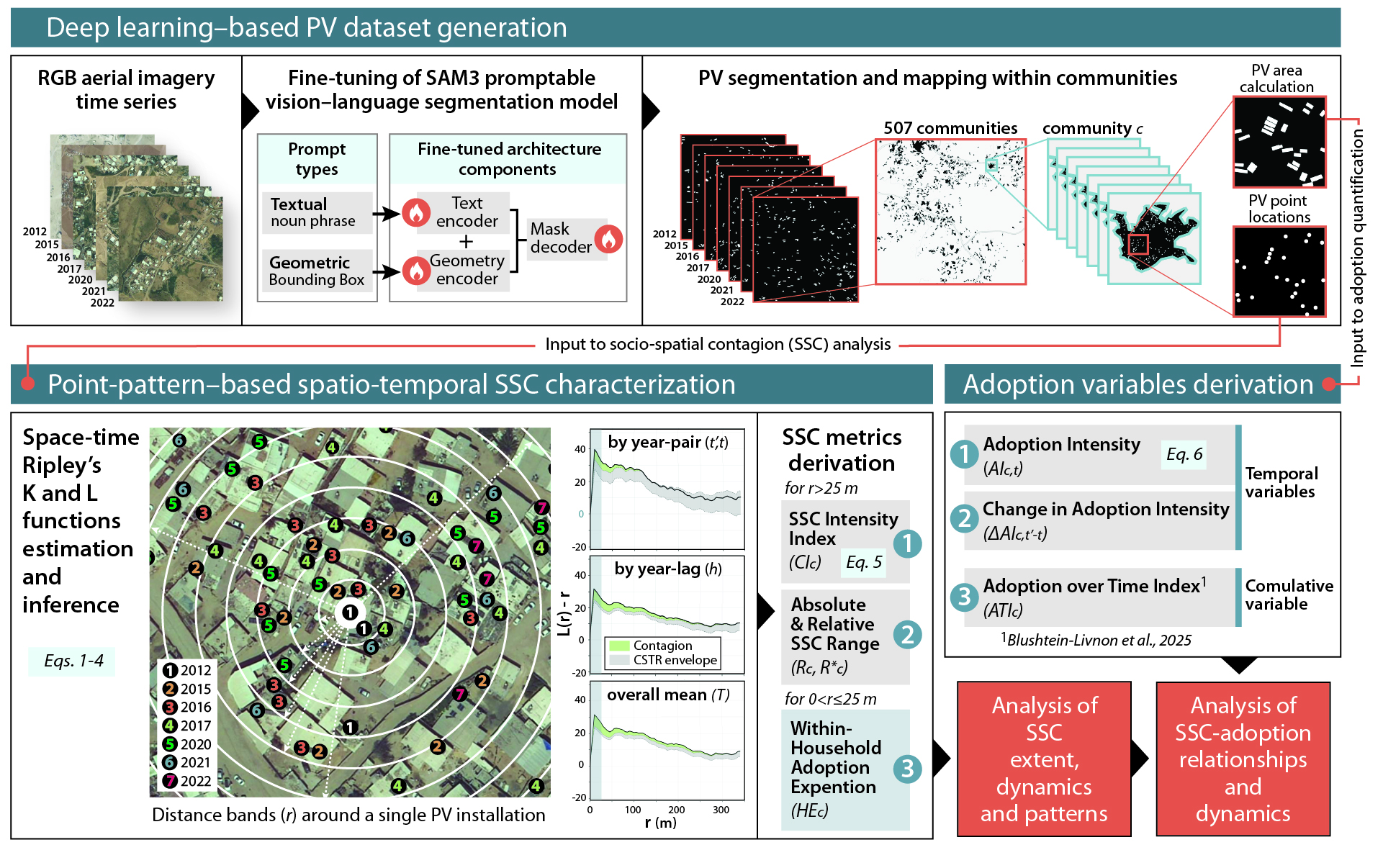}
   \vspace{-20pt}
    \caption{
    \textbf{Overview of research framework.} Aerial imagery was processed using a fine-tuned SAM3 model to segment PV installations over time. The generated data support point-pattern analysis to derive SSC metrics across multiple temporal scales, and to quantify temporal and cumulative adoption intensity. The framework enables integrated analysis of SSC extent, dynamics, and patterns, and their relationships with the adoption process.
    } 
    \label{framework}
\end{figure*}

An overview of the research framework is presented in Fig. \ref{framework}.

\subsection{Study Area}
This study draws on an empirical case from the Negev Desert, Israel, a sparsely populated semiarid region inhabited by $\sim$100,000 Bedouin residents. The Bedouin population constitutes a traditional rural society organized in \textgreater 1,000 small, spatially fragmented, and unauthorized settlement clusters (referred here as communities) that remain disconnected from the national electricity grid \cite{yahel2019conflict, contreras2023}. These communities are structured around extended family clusters, with kinship ties playing a central role in shaping social and spatial organization \cite{yahel2021tribalism}. Residential structures are predominantly composed of lightweight constructions, forming dwelling clusters with relatively homogeneous internal density. Similar to other rural contexts, particularly in Sub-Saharan Africa (e.g.,\cite{danso2020family, jones2022house, nankabirwa2024}), a single household often comprises multiple residential structures rather than a single dwelling unit. Despite variation in built-up area size across communities, residential structure density remains relatively consistent. These are characterized by persistent infrastructure deficits, extremely low socioeconomic conditions, and acute energy poverty \citep{teschner2024energy}.

Until the early 2010s, household electricity supply in the communities relied on diesel generators \cite{teschner2020, kattan2018}. Over the past decade, residential PVs have emerged as the primary alternative, offering a comparatively affordable and more efficient source of electricity \citep{shapira2021energy}. Preliminary data point to a substantial expansion of PV adoption during this period, accompanied by pronounced spatial heterogeneity in adoption intensity, both across and within communities. This trajectory parallels global trends in decentralized solar energy, as reflected in the fourfold increase in installed off-grid PV capacity worldwide, 2014-2023 \citep{IRENA2024Off}, while the observed spatial disparities mirror diffusion patterns documented in other marginalized off-grid contexts, where adoption processes are highly localized and shaped by place-specific and institutional constraints \citep{Aarakitl2021, akter2021, saha2025}.

The Bedouin case reflects structural conditions common to rural off-grid communities across the Global South, particularly in Sub-Saharan Africa and the Middle East. Shared characteristics include dispersed settlement patterns, chronic infrastructure shortages, severe energy deprivation, and adoption barriers linked to poverty, limited formal education, and restricted access to information \citep{shapira2021energy, IRENA_2025}. These parallels are embedded within a broader global challenge, as an estimated 730 million people remained without access to electricity in 2024, nearly 85\% of whom reside in Sub-Saharan Africa \citep{IEA2025Access, ESMAP2024}. Against this backdrop, and given comparable climatic conditions and exposure to climate-related risks such as droughts and flooding \citep{SDG72025}, the Bedouin case provides a theoretically and empirically grounded basis for deriving broader insights into SSC and diffusion dynamics of off-grid solar adoption.

\subsection{Deep learning-based PV dataset generation}
A spatiotemporal dataset of household-level PV installations was curated using high-resolution RGB aerial imagery from the \textit{Authority for Development and Settlement of the Bedouin in the Negev} and the \textit{Survey of Israel}. Seven orthophotos with spatial resolutions of 0.13–0.25 m/pixel were available for 2012, 2015, 2016, 2017, 2020, 2021, and 2022.

PV installations were detected and delineated using automated semantic segmentation with the SAM3 foundation model \citep{carion2025sam3}, a vision-language architecture that enables promptable concept-level segmentation by encoding textual prompts as abstract visual concepts. 

To adapt SAM3 to overhead imagery and target class, comprising rooftop and ground-mounted PV systems, a fine-tuning pipeline integrating semantic and spatial conditioning was applied. A spatial prompt was provided via bounding boxes derived from ground-truth masks, and textual guidance was defined using a target-specific noun phrase.

The textual prompt was selected from a pool of 10 domain-relevant noun phrases derived from Wikidata \citep{vrandevcic2014wikidata}, which is incorporated within SAM3’s SA-Co ontology. Candidate prompts were evaluated on 50 annotated image-mask pairs using zero-shot, text-only prompting and ranked by mean F1 and IoU scores. The highest-scoring phrase was selected as the final prompt.

During fine-tuning, the text and geometry encoders were unfrozen to enable joint semantic–spatial adaptation, and the mask decoder was unfrozen to support output refinement. The vision encoder and DETR-based encoder–decoder were kept frozen to preserve pretrained representations and ensure stable training. Training was conducted using 700 annotated samples, with 200 for validation and 500 for testing, for each time point. Inference was conducted using textual prompting. Performance was evaluated, at the pixel level, on the held-out test set. Model performance across the evaluated metrics by year is reported in Appendix \ref{model}; descriptive statistics of the detected PVs by year are provided in Appendix \ref{panels}.

Pixel-wise segmentation results were converted into object-level representations. For each detected panel, its surface area was calculated from the corresponding object footprint and subsequently used as the basis for computing $\textit{ATI}$ \cite{blushtein2025beyond} and $\textit{AI}$ (Eq. \ref{AI}) metrics. The resulting outputs were then converted to point representations to form a spatially explicit time series of adoption events, serving as input layers for point-pattern analyses (section \ref{ppa}).

\subsection{Spatio-temporal analysis of SSC} \label{ppa}
To quantify the extent and dynamics of SSC in PV adoption, we employed the spatio-temporal Ripley’s $K$ function \citep{lee2023spatio, illian2008}. This class of second-order statistics is widely used to detect departures from Complete Spatio-Temporal Randomness (CSTR) and was applied across diverse fields to analyze diffusion processes and localized interactions in space and time, including epidemiology \citep{hohl2016, yue2020drought}, economics \citep{kosfeld2011spatial}, ecology \citep{hendricks2017}, environmental hazards \citep{wang2020spatiotemporal}, and urban studies \citep{fu2017study}.

Let $(l_i, t_i)$ denote the spatial location and installation time of a PV system $i$, observed within a bounded spatial window $W$ and temporal window $T$. For any ordered pair of installations $(i,j)$, $\Delta l_{ij}$ denotes the Euclidean distance between locations $l_i$ and $l_j$, and $\Delta t_{ij}$ denotes their forward temporal lag, with $t_j > t_i$.

The spatio-temporal $K$ function, $K(r,\tau)$, measures the expected number of additional installations occurring within spatial distance $r$ and temporal lag $\tau$ of an arbitrary installation, normalized by the overall spatio-temporal intensity of the process $\lambda$ (Eq. \ref{ripley1}-\ref{ripley2}). Spatial distances $r$ were evaluated at fixed increments of 10~m.
\vspace{-5pt}
\begin{equation}
\small
K(r,\tau)
=
\frac{1}{\lambda}
\,
\mathbb{E}
\left[
\sum_{j \neq i}
\mathit{I}\,(\Delta l_{ij}, \Delta t_{ij})
\right],
\label{ripley1}
\end{equation}
\vspace{-10pt}
\begin{equation}
\small
\lambda = \frac{N}{|W| \cdot |T|},
\label{ripley2}
\end{equation}
where $N$ is the total number of observed installations, and $\mathit{I}\,(\cdot)$ is the indicator function, defined as
\begin{equation}
\small
\mathit{I}\,(\Delta l_{ij}, \Delta t_{ij}) =
\begin{cases}
1, & \text{if } \Delta l_{ij} \le r \text{ and } 0 < \Delta t_{ij} \le \tau,\\
0, & \text{otherwise}.
\end{cases}
\label{indicator}
\end{equation}

To facilitate interpretation, we further employed the variance-stabilized $L$ function:
\begin{equation}
L(r,\tau) = \sqrt{\frac{K(r,\tau)}{\pi}} - r,
\label{ripley3}
\end{equation}
which linearizes the expected value under a CSTR null model. Under the assumption that PV installations occur independently across space and time, $L(r,\tau)$ is expected to fluctuate around zero for all spatial distances and temporal lags.

\subsubsection{Interpretation and statistical inference}
Positive $L(r, \tau)$ values indicate SSC, implying that more installations occur within distance $r$, and time lag $\tau$, than would be expected by chance. For PV adoption, these values are interpreted as evidence of localized SSC, in which recent nearby installations increase the likelihood of subsequent adoption. Values of $L(r,\tau)$ $\approx$0 indicate that the observed pattern does not depart from CSTR with the same overall event intensity.

Statistical significance was assessed using Monte Carlo simulations under a null model that preserves the overall intensity of installations while randomizing their spatio-temporal arrangement. Simulation envelopes were constructed from 1,000 repeated realizations of the null process, and observed values of $L(r, \tau)$ falling outside these envelopes were interpreted as statistically significant departures from CSTR.

Two interpretive clarifications are central for understanding Ripley’s spatio-temporal function. \textit{First}, statistically significant positive values of $L(r,\tau)$ do not imply higher adoption or a greater number of installations. Rather, they indicate that installations occurring within the examined temporal lag are distributed non-randomly in space and time, being significantly more likely to emerge near prior installations within the specified temporal lag. Thus, evidence of SSC captures the spatio-temporal organization of newly added installations relative to existing ones, and not the adoption growth magnitude. \textit{Second}, the persistence of positive $L(r,\tau)$ values across a range of spatial distances and temporal lags signals that SSC is not confined to a single scale, but rather reflects sustained local dependence in spatial placement and timing of events, and does not imply homogeneous spatial density.

\subsubsection{Geometric constraints and boundary effects}
To ensure adequate statistical power, the analysis was restricted to communities with $\ge$50 observed PV installations across the full study period, yielding a final sample of 507 communities. Built-up areas across communities exhibit substantial heterogeneity in both size and shape, with areas varying by nearly an order of magnitude, ranging from approximately 7,000 to 66,000~m$^{2}$, with direct implications for distance specification. Accordingly, a community-specific maximum spatial distance $r_{\max}$ was defined for each community rather than applying a single global cutoff, enabling normalization of spatial distances relative to each community’s spatial extent and ensuring comparability across communities.

$r_{\max}$ was derived from the geometry of each community polygon, using an area-based reference radius $r_c = \sqrt{A_c/\pi}$, where $A_c$ is the polygon area, along with an additional compactness adjustment based on the polygon’s bounding-box eccentricity. For near-compact polygons (bounding-box aspect ratio$\,\leq$$1.5$), $r_{\max}$ was set to $r_c$, whereas for elongated polygons it was constrained to $\min(r_c,\, 0.8\,r_{\max})$, where $r_{\max}$ denotes the maximum chord length estimated from the polygon boundary coordinates.

Edge correction was applied during estimation to account for truncated space-time cylinders associated with events near the observation window boundaries. To further reduce boundary-related bias, the effective analysis distance was set to $0.9\,r_{\max}$, thereby excluding distances approaching the polygon boundary, where edge effects become dominant.

\subsection{Variables}
\subsubsection{SSC intensity index}
For each community and each year pair $(t,t')$ with $t'>t$, we summarize the corresponding $L$ curve using a normalized SSC intensity index (denoted as $\textit{CI}$), designed to capture the magnitude and the statistical significance of SSC. Let $L(r)$ denote the observed $L$ curve, and let $L^{+}(r)$ and $L^{-}(r)$ denote the upper and lower bounds of the global simulation envelope.

$\textit{CI}$ is defined as the normalized area by which the observed curve exceeds the upper envelope (Eq. \ref{CI}):
\begin{equation}
\textit{{CI}}_{t,t'}
=
\frac{
\displaystyle
\int_{0}^{r_{\max}}
\big[L_{t,t'}(r) - L^{+}_{t,t'}(r)\big]_{+} \, dr
}{
\displaystyle
\int_{0}^{r_{\max}}
\big[L^{+}_{t,t'}(r) - L^{-}_{t,t'}(r)\big] \, dr
}.
\label{CI}
\end{equation}
\noindent
Here, $[x]_{+} = \max(x,0)$.

The numerator of $\textit{CI}_{t,t'}$ represents the net SSC, beyond what would be expected under the null model, integrated over distance. The denominator corresponds to the area of the global simulation envelope and serves as a normalization term that scales the departure relative to the overall uncertainty of the null distribution. This normalization renders $\textit{CI}$ dimensionless and ensures comparability across communities with different spatial extents, event densities, and envelope widths.

$\textit{CI}$ was computed and analyzed at three temporal aggregation levels. 
At the finest level, $\textit{CI}_{t,t'}$ values were calculated for all ordered year pairs. 
At an intermediate level, these pairwise indices were grouped by calendar lag $h$=$t'$$-t$, measured in elapsed years, and averaged within each lag to obtain lag-specific $\textit{CI}$, denoted $\overline{\textit{CI}}$${_h}$, with $h$$\,\in$$\{1,..., 5\}$. 
At the most aggregated level, a community-level summary index was computed by averaging $\overline{\textit{CI}}$${_h}$ across all available lags. This final index, denoted $\overline{\textit{CI}}$, captures the overall intensity of SSC across all communities over the entire observation period.

\subsubsection{Absolute and relative SSC range}
We quantified the range over which statistically significant SSC is detected for each community using both absolute and relative measures. The absolute SSC range, $R$, captures the maximum spatial extent of SSC in meters and is defined by aggregating all distances at which the observed $L$ curve exceeds the upper simulation envelope. To facilitate comparison across communities of different sizes, a relative range measure was defined as $R^{*}$=$ R/r_{\max}$, normalizing distances by each community’s maximum spatial extent.

SSC ranges were computed and analyzed at three temporal aggregation levels. At the finest level, absolute and relative ranges were calculated for each community and ordered year pair. At an intermediate level, these pairwise range measures were grouped by calendar lag and averaged within each lag to obtain lag-specific ranges. At the most aggregated level, lag-specific ranges were further averaged across all available lags, within each community, yielding a descriptor of the typical spatial reach of SSC over the study period.

\subsubsection{Adoption intensity}
Adoption intensity measures the spatial density of PV installations per community, computed as total PV surface area ($PVa$) divided by built-up area ($Ba$). Normalization by area rather than household count is used due to a lack of demographic data. Adoption intensity $AI_{c,t}$ was computed for each community $c$, at time $t$ and defined as (Eq. \ref{AI}):
\vspace{-2pt}
\begin{equation}
\textit{AI}_{c,t} = \frac{PVa_{c,t} \cdot 10^6}{Ba_c}.
\label{AI}
\end{equation}
The scaling constant $10^6$ is applied to improve numerical stability and interpretability.

\subsubsection{Adoption over time index}
The Adoption over Time Index ($\textit{ATI}$) is a composite index, developed by us, to capture cumulative adoption intensity within a given community over the entire observation period through temporal integration. By integrating adoption intensity over time, $\textit{ATI}$ jointly reflects both the overall magnitude of adoption and its temporal trajectory, including fluctuations in growth rate. $\textit{ATI}$ values are normalized relative to the regional mean, enabling direct comparison of each community’s adoption performance with the regional benchmark. A detailed description of the index and its rationale is provided in \citet{blushtein2025beyond}.

\subsubsection{Within-household adoption expansion}
In addition to SSC across households, we explicitly distinguish within-household expansion in PV adoption, a separate process that captures a household's accumulation of PV installations over time. In many off-grid contexts, PV adoption often proceeds incrementally rather than as a single installation event \citep{GOGLA2023}, as households expand system capacity in response to changing needs, gradual accumulation of trust in technology, and evolving access to resources \citep{lemaire2018solar, mahieu2025, GOGLA2023}.

This distinction is relevant in the Bedouin case, as in other traditional rural contexts in the Global South, where a household typically comprises multiple spatially proximate structures, including auxiliary and guest buildings, as well as separate dwellings for wives and their children in polygynous households. Consequently, installations associated with a single household are frequently distributed across several nearby structures, yielding an average household spatial footprint of $\sim$50m in diameter. Accordingly, distances up to 25m (the approximate radius of this footprint) are attributed to within-household interactions, providing an empirically grounded approximation of intra-household spatial extent. Within-household adoption expansion ($\textit{HE}$) was therefore quantified using the same Ripley’s $L$ framework, with values evaluated over spatial lags $r \in [0, 25]$ interpreted as within-household expansion rather than cross-household SSC.

\subsection{SSC patterns} \label{patterns}
To characterize heterogeneity in SSC regimes, communities were classified into four patterns based on two conceptual dimensions: intensity and range. The first dimension is operationalized using the SSC intensity Index ($\textit{CI}$). The second dimension represents the spatial extent over which SSC operates and is quantified by absolute SSC range ($R$) and relative SSC range ($R^*$). Using a relative measure accounts for variation in community size, preventing small communities with extensive SSC relative to their size from being misclassified as short-range.

Both dimensions were dichotomized using mean-based thresholds, providing a transparent and interpretable benchmark for distinguishing above-average SSC levels across communities. While alternative thresholding strategies (e.g., quantile-based) can also be applied, the mean offers a simple, reproducible criterion. Intensity was classified as high when $\textit{CI}$>1.15 and low otherwise. Range was classified as long-range when either $R$>105.8 m or $R^*$>0.86, and as short-range only when both $R\leq$105.8 m and $R^*\leq$0.86.  Combining the dual dimensions of SSC yields four distinct patterns: high-long, high-short, low-long, and low-short.

\subsection{Transitions in SSC patterns and adoption dynamics}
To address the third objective, we examined how SSC patterns evolve over the adoption process and how these dynamics relate to changes in adoption intensity over time. 
Specifically, we ask: (1) whether communities exhibit systematic transitions between SSC patterns over time; (2) whether transitions between SSC patterns are associated with systematic differences in changes in adoption intensity; and (3) whether this association varies over the course of the adoption process. Changes in SSC patterns are examined separately along the two core dimensions of SSC identified here - SSC intensity and range.

Given that SSC patterns are defined over year pairs $(t,t')$, their dynamics are examined across three transition windows, each defined by consecutive year pairs, as summarized in Table~\ref{transition_windows}.

\begin{table}[t]
\centering
\footnotesize
\caption{Transition windows across the observed timeline}
\label{transition_windows}

{\renewcommand{\arraystretch}{1.15}
\begin{tabular}{%
L{0.16\columnwidth}
C{0.18\columnwidth}
C{0.22\columnwidth}
C{0.22\columnwidth}
}
\toprule
\textbf{Phase} &
\textbf{\makecell{Transition\\[-2pt] window}} &
\textbf{\makecell{Initial\\[-2pt] year pair}} &
\textbf{\makecell{Final\\[-2pt] year pair}} \\
\midrule
Early        & $T_1$ & 2012-2015 & 2015-2017 \\
Intermediate & $T_2$ & 2015-2017 & 2017-2020 \\
Late         & $T_3$ & 2017-2020 & 2020-2022 \\
\bottomrule
\end{tabular}
}
\end{table}

Consistent with the classification framework (Section~\ref{patterns}), SSC patterns were defined along the same two dimensions, but evaluated separately within each year-pair window. For each year pair $(t,t')$, mean SSC intensity and range were computed, and pattern thresholds were set relative to the contemporaneous distribution, so pattern assignments reflect each community’s position within period-specific thresholds. 

Based on these period-specific assignments, changes in SSC intensity and range were identified by comparing each community’s classification across two consecutive year pairs. Transitions were classified as \textit{upward} when shifting from low/short to high/long SSC intensity/range, \textit{downward} when shifting from high/long to low/short categories, and \textit{stable} when no categorical shift occurred. Stability thus indicates persistence within the same SSC pattern rather than constant absolute SSC levels, as mean intensity and range of SSC evolve over time. Upward and downward transitions capture shifts relative to period-specific mean thresholds. 
Accordingly, observed transitions reflect relative strengthening or weakening of SSC dynamics over time against a background of generally increasing absolute levels.

To address the \textit{first} question, we employed multinomial logit models (MLM) \cite{kwak2002multinomial} for each SSC dimension to test whether the likelihood of different transition types varies systematically across transition windows. 

To address the \textit{second} and \textit{third} questions, we estimate separate linear regression models for each SSC dimension, relating transitions in SSC patterns to changes in adoption intensity over time. Let $\Delta{\textit{AI}_{c,k}}$ denote the dependent variable, defined as the relative change in adoption intensity of community $c$ over transition window $k$, with $k \in \{T_1, T_2, T_3\}$. The relative change is computed with respect to the baseline adoption intensity at the beginning of the transition window. To accommodate zero baseline values, a constant of 1 is added to the baseline term in the denominator, ensuring that the measure remains well-defined and comparable across communities.

Changes in SSC patterns are represented using indicator variables for upward $(U)$ and downward $(D)$ transitions, with stable transitions serving as the reference category. The models relate $\Delta{\textit{AI}_{c,k}}$ to transition type, transition window, and their interaction, as specified in Eq.~\ref{reg_model}.

Joint Wald tests are used to assess both the overall association between SSC pattern transitions and relative changes in adoption intensity, and the extent to which this association varies across transition windows. All models are estimated with community-level cluster-robust standard errors to account for repeated observations of the same communities across transitions.
\vspace{-8pt}
\begin{equation}
\resizebox{\columnwidth}{!}{$
\Delta\textit{AI}_{c,k}
=
\alpha
+
\beta_U \, U_{c,k}
+
\beta_D \, D_{c,k}
+
\gamma_k
+
\delta_{U,k} \, U_{c,k}
+
\delta_{D,k} \, D_{c,k}
+
\varepsilon_{c,k}
$}
\label{reg_model}
\end{equation}
The term $\gamma_k$ captures overall temporal differences in changes in adoption intensity across transition windows. 
The interaction terms $\delta_{U,k}$ and $\delta_{D,k}$ allow the association between upward and downward SSC pattern transitions and changes in adoption intensity to vary across transition windows.

\section{Results}
\subsection{SSC extent and dynamics}
\vspace{2pt}
\subsubsection{SSC existence and extent}
Across 507 communities, SSC in PV adoption is nearly ubiquitous. Only $\sim$5\% of the communities exhibit no statistically detectable SSC, with community-level mean Ripley’s $L$ curve remaining entirely within simulation envelopes for spatial distances of $\mathord{>}25\, m$, $\overline{\textit{CI}}$\,$\approx$\,0. 
The mean SSC intensity across communities indicates a generally high level of clustering ($\overline{\textit{CI}}$=1.15; Table~\ref{descriptive}). Given the definition of $\textit{CI}$ as a normalized integral (Eq.~\ref{CI}), this reflects that the cumulative area by which $L(r)$ exceeds the upper envelope, exceeds the total area of the simulation envelope under CSTR by a factor of 1.15. The distribution of $\overline{\textit{CI}}$ is right-skewed, with the mean exceeding the median and values reaching up to $\overline{\textit{CI}}$=3.57. Variability is considerable (SD=0.65; IQR=0.78), pointing to substantial heterogeneity in spatio-temporal clustering.

\begin{table}[b!]
\centering
\footnotesize
\caption{Descriptive statistics of SSC intensity and range across communities}
\label{descriptive}
\renewcommand{\arraystretch}{1.05} 
\setlength{\extrarowheight}{0.2pt} 
\setlength{\tabcolsep}{18pt} 
\begin{tabular*}{\columnwidth} {@{\hspace{10pt}} p{0.2\columnwidth} p{0.07\columnwidth} p{0.1\columnwidth} p{0.12\columnwidth} @{\hspace{6pt}}} \toprule
Statistic & $\mathbf{\overline{CI}}$ & $\mathbf{R}$\hspace{0.6pt}{\scriptsize(m)} & $\mathbf{R^*}$\hspace{-0.2pt}{\scriptsize (\%)} \\
\addlinespace[-0.5pt] 
\midrule
Mean       & 1.15  & 105.80 & 86.24  \\ 
SD         & 0.65  & 49.66 & 10.60  \\
Median     & 1.07  & 101.02 & 90.02  \\
Q1 {\scriptsize (25\%)}  & 0.69  & 70.02 & 82.23  \\
Q3 {\scriptsize (75\%)}  & 1.47  & 129.90 & 92.30 \\
Min        & 0.00  & 25.00 & 11.79  \\
Max        & 3.57  & 400.00 & 97.61 \\
\bottomrule
\end{tabular*}
\par\vspace{3pt}
\scriptsize
\raggedright
\textbf{Notes:} Range statistics are computed for distances >25 m (to exclude within-household expansion).
\end{table}

SSC range is similarly extensive, with a mean absolute range of $\sim$106~m and a mean relative range of $\sim$86\%. While absolute SSC ranges vary widely and are strongly influenced by community built-up area size, relative SSC ranges are more tightly distributed, with a median of 90\% and an IQR of $\sim$10\%. This suggests that SSC typically operates at the scale of the entire community rather than being confined to immediate neighbors.

The results described above demonstrate that SSC in PV adoption is both widespread and spatially extensive in off-grid communities, directly addressing the first component of Objective 1 by establishing its presence and quantifying its typical spatial extent, which often extends across substantial portions of the community. The following subsection examines how the strength and spatial range of SSC evolve over time.

\subsubsection{SSC dynamics}
\begin{figure*}[t]
    \centering
    \includegraphics[width=0.95\textwidth]{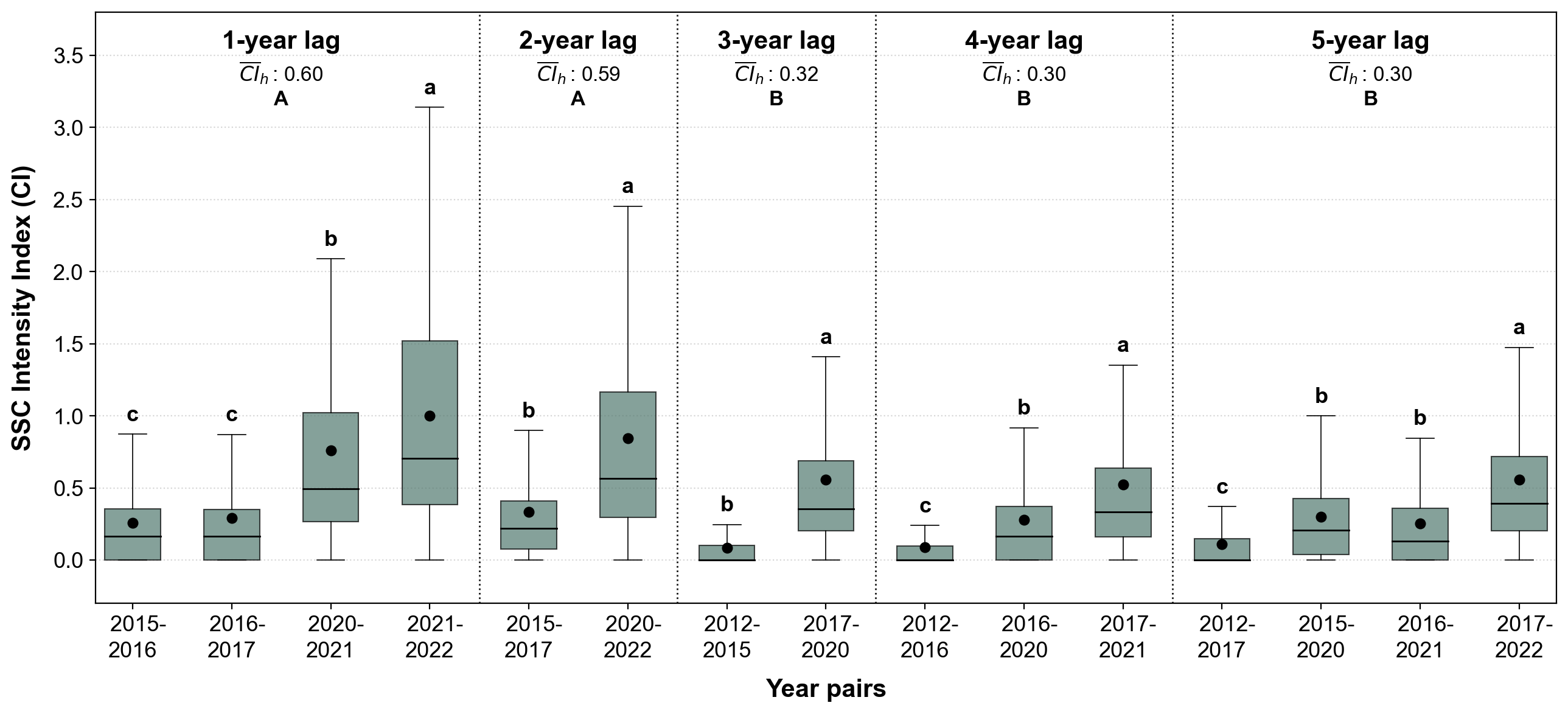}
    \vspace{-5pt}
    \caption{
    \textbf{SSC intensity dynamics across temporal lags.} $\textit{CI}$ increases over time within each lag and is strongest at short temporal lags of one to two years, declining at longer lags. Lowercase letters denote significant differences (p<0.001) between year pairs within each temporal lag; Uppercase letters indicate significant differences in mean $\textit{CI}$ across temporal lags ($\overline{\textit{CI}}$${_h}$). Black dots represent the mean $\textit{CI}$ for each year pair ($\overline{\textit{CI}}_{t,t'}$).}
    \label{CI-lags}
\end{figure*}

Figure~\ref{CI-lags} presents the dynamics of $\textit{CI}$ across different year lags, where each lag represents the elapsed time between installation events, with results shown for successive year-pairs corresponding to each lag. Within each lag group, $\textit{CI}$ shows a clear upward trend across later-year pairs, indicating a systematic strengthening of SSC over time. One-way ANOVA indicates statistically significant differences among year pairs within each lag (p<0.001), and Tukey post hoc tests show that most pairwise contrasts are significant. Alongside rising central tendency over time, $\textit{CI}$ also shows increasing dispersion, with later periods at each lag exhibiting broader distributions and more pronounced upper-end values, indicating growing heterogeneity in SSC intensity across communities.

\begin{figure*}[t]
    \centering
    \includegraphics[width=0.95\textwidth]{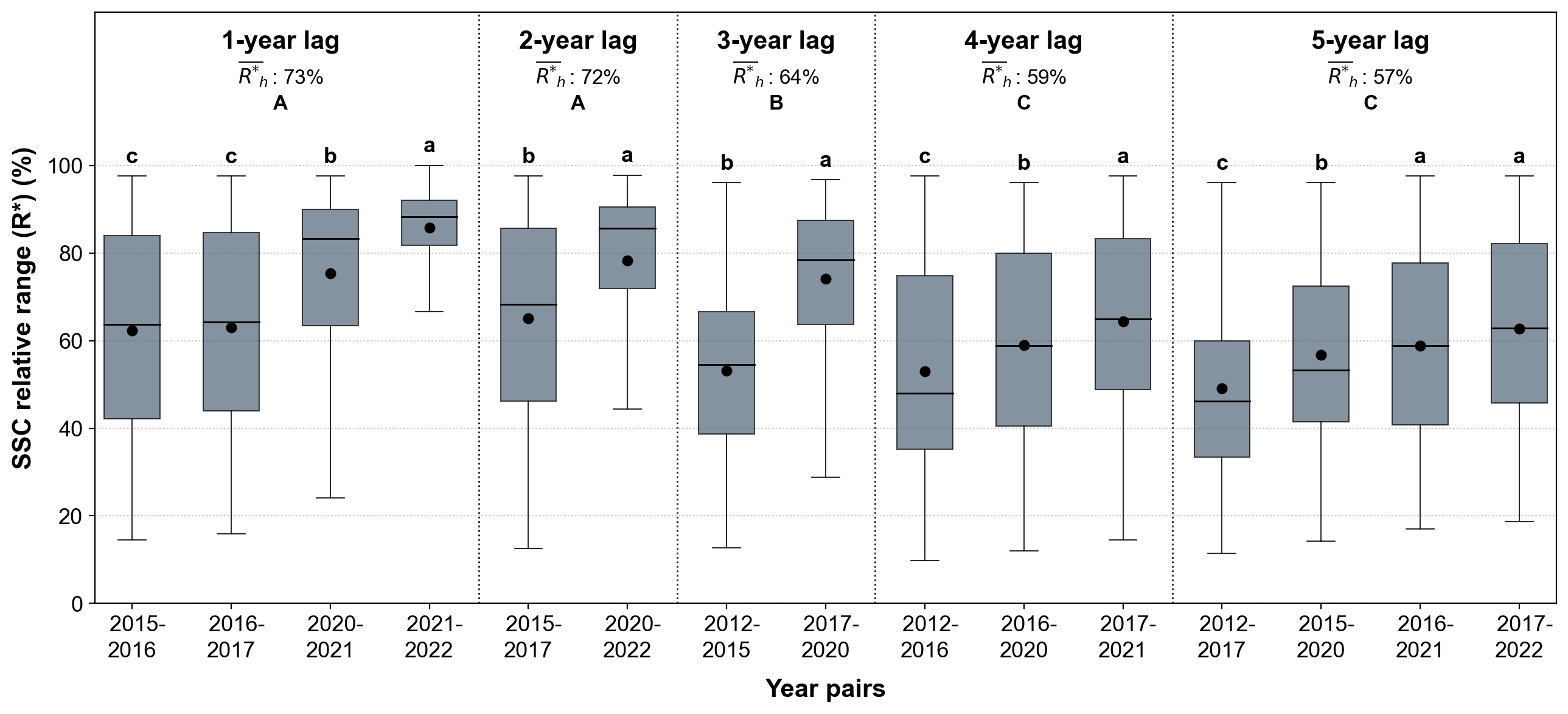}
    \vspace{-8pt}
    \caption{
    \textbf{SSC relative range dynamics across temporal lags.} 
The mean relative range of SSC ($\overline{\textit{R}\,^*}\!{_h}$) increases over time within each temporal lag, while dispersion across communities decreases. Across temporal lags, the SSC range exhibits a three-tier structure, declining monotonically with increasing lag. Lowercase and uppercase letters denote significant differences (p<0.001) between year pairs within temporal lags and across temporal lags, respectively. Black dots represent the mean SSC range for each year pair.
    }
    \label{RR-lags}
\end{figure*}

\begin{figure*}[!htbp]
    \centering
    \includegraphics[width=0.95\textwidth]{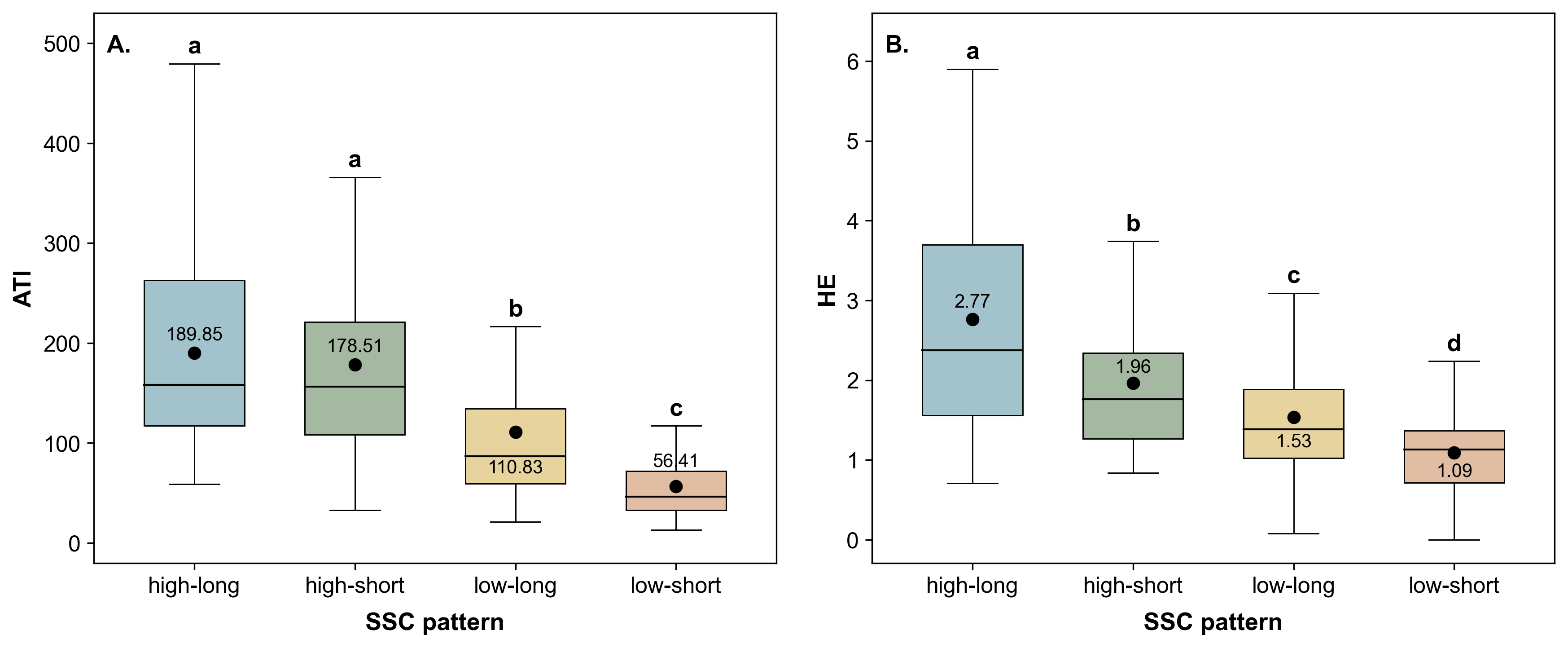}
    \vspace{-4pt}
    \caption{
    \textbf{Adoption outcomes across SSC patterns.} A. SSC patterns by adoption over time index ($\textit{ATI}$); B. SSC patterns by within-household adoption expansion ($\textit{HE}$). Higher SSC intensity is associated with higher adoption intensities. Letters denote significant differences between patterns (p<0.001); black dots indicate the mean value.
    }
    \label{ATI_HE}
\end{figure*}

Across lag groups, a strong temporal decay trend is evident. Lag-specific mean $\textit{CI}$ values are significantly higher for one- and two-year temporal lags, whereas both differ significantly from three- to five-year lags, which form a lower, statistically indistinguishable group. This trend indicates that SSC is most pronounced over short temporal windows, in the immediate aftermath of nearby installations, and attenuates sharply as the temporal separation between installations increases.

Figure \ref{RR-lags} depicts the dynamics of the spatial reach of SSC, measured by relative SSC range, across year pairs grouped by year lag. Within each lag, the SSC range increases over time, accompanied by systematic convergence in the distribution, indicating greater homogeneity in the relative range as adoption progresses. Differences between year pairs within each lag are significant in nearly all cases (p<0.001), confirming a robust temporal ordering of SSC range dynamics.

Across year lags, mean relative ranges of SSC clustered into three statistically distinct groups. One- and two-year lags exhibit a similarly high range, with no significant difference between them. The three-year lag forms an intermediate group, while four- and five-year lags show significantly narrower relative ranges. Together, these dynamics indicate that SSC increases as adoption progresses, operates over its broadest and most consistent distances shortly after nearby installations, and becomes increasingly spatially constrained as temporal separation increases.

These results establish SSC as a pervasive and structurally embedded feature of PV adoption in off-grid communities. SSC is not only widespread across communities but also spans substantial spatial distances that typically encompass much of the community footprint, while remaining temporally bounded within relatively short windows following prior installations. By jointly documenting its prevalence, spatial reach, and temporal decay, this analysis addresses Objective 1 by providing a comprehensive characterization of the existence, extent, and dynamics of SSC in off-grid PV adoption.

\subsection{SSC patterns and adoption outcomes}
While the preceding analyses document the existence, extent, and dynamics of SSC, they do not capture how these dimensions combine into distinct spatial patterns at the community level, nor do they examine their association with adoption outcomes. We therefore classify communities into four SSC patterns defined by the joint configuration of SSC intensity and range. 

Table~\ref{patterns_desc} summarizes the defining characteristics of the four SSC patterns. The distribution of communities across patterns is uneven, with high-intensity SSC accounting for nearly half of the sample. Long-range SSC is more prevalent than short-range SSC overall, accounting for 62\% of communities.

Figure~\ref{ATI_HE} links the four SSC patterns to adoption outcomes, captured by two variables: the \textit{adoption over time index} ($\textit{ATI}$) and \textit{within-household adoption expansion} ($\textit{HE}$). As $\textit{CI}$ measures spatial clustering rather than adoption magnitude or growth, these variables provide complementary indicators of adoption performance across SSC patterns.

Statistically significant differences emerge across SSC patterns for both $\textit{ATI}$ and $\textit{HE}$ (one-way ANOVA with Tukey post hoc tests, p<0.001). Communities characterized by high-intensity SSC exhibit substantially higher cumulative adoption, irrespective of spatial range (as the high-long and high-short patterns show similar $\textit{ATI}$ values), whereas both low-intensity patterns display markedly lower $\textit{ATI}$ values, with a substantial decline from long to short SSC range.

A parallel, yet more differentiated, trend is observed for $\textit{HE}$. High-intensity SSC is associated with significantly higher $\textit{HE}$ values, indicating a more pronounced accumulation of PV installations within households over time. Unlike $\textit{ATI}$, spatial range further stratifies $\textit{HE}$ within high-intensity SSC. Communities in the high-long pattern exhibit the highest $\textit{HE}$ levels, followed by high-short, while both low-intensity groups show limited expansion, with the lowest values recorded in low-short communities.

These results address Objective 2 by showing that communities with higher SSC intensity consistently exhibit greater cumulative adoption and more pronounced within-household expansion over time. Spatial range further differentiates patterns of within-household expansion even among communities with similarly high overall adoption intensity ($\textit{ATI}$). Within-household expansion systematically covaries with SSC range across neighboring households, indicating that SSC dynamics and adoption extent evolve in tandem throughout the diffusion process.

\begin{table}[t]
\centering
\footnotesize
\caption{Descriptive statistics of SSC patterns}
\label{patterns_desc}
\renewcommand{\arraystretch}{1.0}
\setlength{\tabcolsep}{3.5pt}
\newcommand{\StatLabelW}{1.3em}

\begin{tabularx}{\columnwidth}{L{0.12\columnwidth} Y Y Y Y}
\toprule
 & \textbf{high-long} & \textbf{high-short} & \textbf{low-long} & \textbf{low-short} \\
\midrule
\%\hspace{1.3pt}{\scriptsize($N$)} &
33.4\hspace{1.3pt}{\scriptsize (169)} &
16.9\hspace{1.3pt}{\scriptsize (86)} &
28.6\hspace{1.3pt}{\scriptsize (145)} &
21.1\hspace{1.3pt}{\scriptsize (107)} \\
\midrule
$\overline{\textbf{\textit{CI}}}$\hspace{4pt}\makebox[\StatLabelW][l]{{\scriptsize Mean}} &
1.58 & 1.62 & 0.71 & 0.61 \\
\hspace*{\StatLabelW} {\scriptsize SD} &
0.55 & 0.57 & 0.24 & 0.31 \\
\addlinespace[2pt]
$\textit{\textbf{R}}$\hspace{8pt}\makebox[\StatLabelW][l]{{\scriptsize Mean}} &
127.91\hspace{1pt}{\scriptsize m} &
58.37\hspace{1pt}{\scriptsize m} &
138.31\hspace{1pt}{\scriptsize m} &
66.90\hspace{1pt}{\scriptsize m} \\
\hspace*{\StatLabelW} {\scriptsize SD} &
45.03\hspace{1pt}{\scriptsize m} &
10.61\hspace{1pt}{\scriptsize m} &
45.58\hspace{1pt}{\scriptsize m} &
27.11\hspace{1pt}{\scriptsize m} \\
\addlinespace[2pt]
$\textit{\textbf{R}}\,^*$\hspace{4pt}\makebox[\StatLabelW][l]{{\scriptsize Mean}} &
92.0\% & 79.0\% & 91.0\% & 71.0\% \\
\hspace*{\StatLabelW}\hspace{4pt}{\scriptsize SD} &
2.0\% & 6.0\% & 19.0\% & 25.0\% \\
\bottomrule
\end{tabularx}
\end{table}

\subsection{Transitions in SSC patterns}
\vspace{3pt}
\subsubsection{Transitions across the adoption process}
\begin{figure*}[H!]
    \centering
    \includegraphics[width=0.95\textwidth]{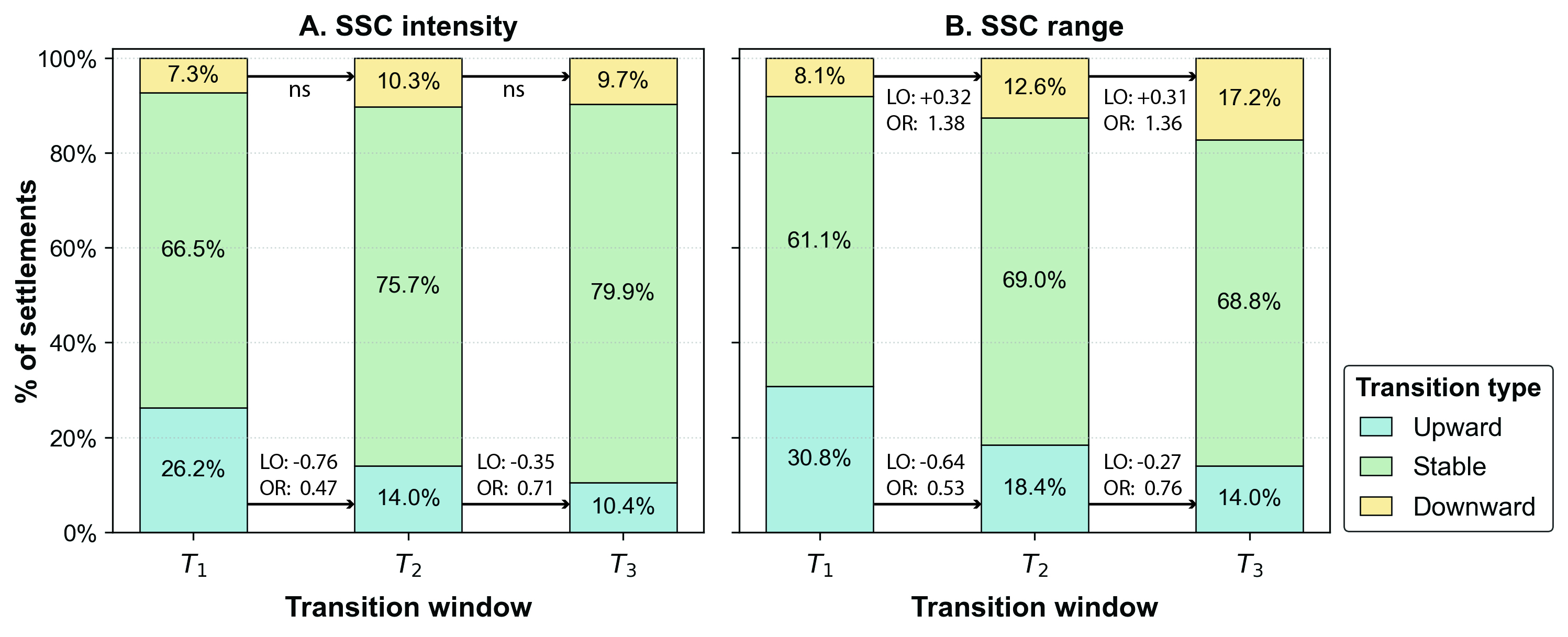}
    \vspace{-5pt}
    \caption{
    \textbf{Transitions in SSC patterns.} Distribution of communities by transition type across three transition windows is shown for SSC intensity (A) and SSC range (B). Stability dominates across all periods, while upward transitions are more prevalent in early phases. MLM estimates shown adjacent to the arrows report log-odds (LO) and corresponding odds ratios (OR) between successive transition windows; all effects are significant (p<0.001).
    } 
    \label{transitions}
\end{figure*}

Figure~\ref{transitions} summarizes the distribution of communities across transition types between consecutive periods, separately for SSC intensity and spatial range. Across both dimensions, stability is the dominant transition type in all periods, indicating that most communities retain their classification relative to rising mean thresholds over time. Nevertheless, non-stable transitions are substantial, particularly during the early phase of the adoption process.

For SSC intensity, upward transitions are most common in $T_1$ and decline sharply thereafter, while stability increases monotonically from $T_1$ to $T_3$, indicating progressive consolidation of SSC intensity. Downward transitions remain infrequent across all periods. 

A similar yet more volatile dynamic is observed for SSC range. Upward transitions are most prevalent in $T_1$ and decline in $T_2$-$T_3$ as stability increases. However, unlike SSC intensity, downward transitions rise over time, peaking in $T_3$ and indicating progressive spatial contraction.

These descriptive dynamics are further supported by MLM estimates. Relative to $T_1$, the odds of upward transitions decline sharply over time in both SSC dimensions, corresponding to odds ratios of 0.47 at $T_2$, 0.71 at $T_3$ for intensity, and 0.53 at $T_2$ and 0.76 at $T_3$ for range. By contrast, the odds of downward transition in SSC range increase in later periods, corresponding to odds ratios of 1.38 at $T_2$ and 1.36 at $T_3$. All reported effects are statistically significant (p<0.01).

These findings address the descriptive component of Objective 3 by showing that SSC patterns are not static, but evolve systematically over time, with early phases characterized by expansion in SSC dimensions and later phases marked by consolidation and spatial contraction. The implications of these transitions for changes in adoption intensity are examined in the following subsection.

\subsubsection{Transitions and adoption intensity change}
Table~\ref{regression} reports the estimated associations between SSC pattern transitions and changes in adoption intensity ($\Delta{\textit{AI}}$), separately for each SSC dimension (intensity and range). Coefficients are estimated relative to the reference group, namely, stable transitions in the first period ($T_1$). Main effects capture differences in the relative change in $\Delta{\textit{AI}}$ between transition types compared to the reference group. Time effects capture period-specific differences in the relative change in adoption intensity ($\Delta{\textit{AI}}$) that are common to all transition types, and interaction terms indicate how these differences vary across transition types over time.

In $T_1$, upward transitions in SSC intensity were associated with significantly greater adoption intensity gains ($\sim$+7.6$\Delta\textit{AI}$) compared to stable transitions, while downward transitions corresponded to a pronounced and statistically significant decline ($\sim$$-$9.9$\Delta\textit{AI}$). A significant positive interaction is observed for upward transitions in $T_3$, whereas none is observed in $T_2$, indicating that this association becomes substantially stronger, but only later in the adoption process. Specifically, within $T_3$, upward transitions are associated with an additional increase of $\sim$+15.1$\Delta\textit{AI}$ (computed as $7.6 + 7.52$) relative to stable transitions within the same period.

Time-effect estimates indicate significantly larger relative increases in adoption intensity in the later period ($T_3$) than in the early phase ($T_1$), independent of transition type, reflecting the expected ongoing progression of the adoption process. 

Consistent with the intensity dimension, downward transitions in range are significantly associated with reductions in adoption intensities (${\sim}$${-}$8.8$\Delta\textit{AI}$). However, no corresponding significant association was observed for upward transitions.

As adoption advances, the relationship between downward transitions in the SSC range and changes in adoption intensity reverses. While no significant interaction is observed in $T_2$, a significant positive interaction emerges in $T_3$, indicating that communities experiencing a relative-to-the-mean reduction in SSC range exhibit significantly greater relative increases in adoption intensity (${\sim}$${+}$10.1$\Delta\textit{AI}$, computed as $-8.8 + 18.9$) than stable communities. No significant interaction effects are observed for upward transitions.

To further assess whether late-phase SSC reconfigurations are associated with structurally different cumulative adoption outcomes, we compared $\textit{ATI}$ values between communities that experienced downward transitions in SSC range and those that experienced upward transitions in SSC intensity during $T_3$. No statistically significant difference was observed between the two groups (Mann-Whitney test, p=0.15), suggesting that although these transition types are associated with distinct short-term adoption dynamics, they may nonetheless correspond to similar cumulative adoption levels at advanced phases of the diffusion process. These results suggest that such communities occupy broadly comparable positions along the adoption trajectory, rather than reflecting clearly separated stages of adoption.

These results address Objective 3 by showing that transitions in SSC patterns are systematically associated with changes in adoption intensity, and that this association differs between SSC intensity and SSC range over the course of the adoption process. In particular, the results indicate that adoption growth in later phases can coincide with both a stagnation in the range of SSC and an increase in SSC intensity. This combination suggests a reconfiguration of SSC toward shorter, more localized spatial scales, whereby adoption continues to intensify through fine-scale neighborhood reinforcement alongside spatial densification at the community level.

\begin{table}[H]
\centering
\begin{minipage}{0.95\columnwidth}
\caption{Relative changes in adoption intensity across transition type and transition windows}
\label{regression}
\footnotesize
\renewcommand{\arraystretch}{1}
\setlength{\tabcolsep}{6pt}

\newcommand{\se}[1]{{\scriptsize (#1)}}

\begin{tabular*}{\columnwidth}{
@{\extracolsep{\fill}}
@{\hspace{22pt}}p{0.17\columnwidth}
>{\centering\arraybackslash}p{0.30\columnwidth}
@{\extracolsep{1pt}}
>{\centering\arraybackslash}p{0.35\columnwidth}
@{}
}
\toprule
& \makecell{\textbf{SSC}\\\textbf{intensity}}
& \makecell{\textbf{SSC}\\\textbf{range}} \\
\midrule
\addlinespace[2pt]
\multicolumn{3}{l}{\hspace{10pt}\textbf{\textit{Main effects}}} \\
\addlinespace[-1pt]
\cmidrule{1-3}
\addlinespace[1pt]

Intercept &
8.85$^{***}$ &
10.39$^{*}$ \\
{\scriptsize (Stable, $T_1$)} &
\se{2.01} & \se{4.19} \\
\addlinespace[3pt]

Upward &
7.6$^{*}$ &
4.84 \\
&
\se{3.32} &
\se{5.96} \\
\addlinespace[3pt]

Downward &
-9.87$^{***}$ &
-8.79$^{***}$ \\
&
\se{2.67} &
\se{2.47} \\
\addlinespace[1pt]
\cmidrule{1-3}
\addlinespace[-1pt]
\multicolumn{3}{l}{\hspace{10pt}\textbf{\textit{Time effects}}} \\
\addlinespace[-1pt]
\cmidrule{1-3}
\addlinespace[1pt]

$T_2$ &
10.75 &
7.82 \\
& \se{8.14} & \se{4.05} \\
\addlinespace[3pt]

$T_3$ &
17.22$^{***}$ &
15.06$^{***}$ \\
& \se{4.01} & \se{4.45} \\
\addlinespace[1pt]
\cmidrule{1-3}
\addlinespace[-1pt]
\multicolumn{3}{l}{\hspace{10pt}\textbf{\textit{Interactions}}} \\
\addlinespace[-1pt]
\cmidrule{1-3}
\addlinespace[1pt]

Upward$\times$$T_2$ &
4.66 &
4.97 \\
&
\se{3.14} &
\se{7.05} \\
\addlinespace[3pt]

Downward$\times$$T_2$ &
-15.45 &
7.47 \\
&
\se{10.13} &
\se{5.19} \\
\addlinespace[3pt]

Upward$\times$$T_3$ &
7.52$^{**}$ &
8.13 \\
&
\se{2.73} &
\se{4.97} \\
\addlinespace[3pt]

Downward$\times$$T_3$ &
-5.44 &
18.91$^{***}$ \\
&
\se{3.34} &
\se{3.30} \\
\bottomrule
\end{tabular*}

\vspace{2pt}
\scriptsize
\raggedright
\textbf{Notes:} Dependent variable: $\Delta{\textit{AI}}$ (Relative change in adoption intensity).
Reference categories: \textit{Stable}, $T_1$.
Sample: 1{,}521 community-transitions (507 communities).
Transition-window fixed effects included. Community-level cluster-robust SEs in parentheses.
$^{*}$p<0.05, $^{**}$p<0.01, $^{***}$p<0.001.
\end{minipage}
\end{table}

\section{Discussion}
This study examined the extent and dynamics of SSC in residential PV adoption across off-grid communities, using deep-learning-based segmentation to quantify installation rates and perform point-process inference. The following section discusses the findings with respect to the research objectives.

\subsection{SSC extent and dynamics}
The first objective of this study was to establish whether SSC operates in off-grid residential PV adoption and, if so, to characterize its spatial and temporal properties. The findings indicate that SSC is not a marginal feature of the adoption process, but a pervasive organizing mechanism. The near-ubiquity of statistically significant spatio-temporal clustering across communities suggests that PV adoption in these communities is mediated by locally embedded exposure to prior adopters. In doing so, these findings extend empirical evidence on peer effects in solar technology adoption to off-grid rural contexts and highlight the central role of fine-scale socio-spatial proximity in shaping the diffusion process.

A central finding here is that PV SSC operates at the community scale, rather than being confined to immediately adjacent households. Theoretically, this suggests that the effective unit of peer exposure in off-grid rural communities is not the neighboring household alone, but the broader community as an integrated interaction field \citep{Gu2022Peer, Liu2026}. As routine encounters, kin-based ties, and informal exchange are embedded in the physical layout of these localities, visual exposure to existing installations may further reinforce and extend through community-based social connections \citep{Liu2026, wolske2020peer}.

This finding aligns with prior literature on the diffusion of new technologies in traditional societies, which identifies spatially localized social learning as a central mechanism of adoption \citep{BandieraRasul2006, Gu2022Peer}. Yet, the spatial scale of such influence varies across technologies and often operates at sub-village distances, as shown in studies of agricultural technology adoption \citep{Matuschke2009, Xu2022Mutual}. This contrast suggests that the broader effective scale of contagion observed here is not a general feature of rural technology diffusion, but may instead reflect the distinctive properties of residential PV systems: Unlike many other household technologies, PV installations are highly salient and continuously embedded in the built environment. Other technologies typically exhibit limited or no visual presence in everyday space, ranging from adoption that is not observable, such as the use of improved crops varieties \citep{Matuschke2009}, to adoption that can only be indirectly inferred from production outcomes, e.g., enhanced fertilization practices \citep{Xu2022Mutual}, or adoption that is only weakly visible, such as clean energy heating or cooking appliances \citep{Zhu2022Peer}. By contrast, the visibility of PV installations can accumulate through everyday movement and routine activities across the community, extending exposure beyond the immediate surroundings. Moreover, as PV-equipped homes become recognizable markers of self-sufficiency and access to modern energy services, adoption acquires social recognition that can foster community norms and expectations regarding social acceptance of PV systems \citep{opiyo2019impacts, Liu2026}. Such recognition may not be confined to immediate neighbors but circulated through interactions across the community as a whole.

A comparison with grid-connected PV literature further highlights the contextual nature of SSC. Whereas PV contagion ranges observed in affluent Western settings often extend several hundred meters and, in some cases, exceed 1 kilometer \citep{Min2025, Bollinger2022, Graziano2019}, the absolute distances identified here are considerably shorter. This difference should not be interpreted as indicating weaker SSC. Rather, it reflects the smaller spatial scale at which local networks of relatives, friends, and neighbors operate in these communities, where everyday social interactions tend to unfold within more compact spatial environments \citep{BandieraRasul2006, Conley2010, Wen2021Clean}.

The temporal results of SSC reveal two complementary patterns: (1) contagion peaks within a lag of 1-2 years after prior installations; and (2) it strengthens as the adoption process progresses. The first observation is broadly consistent with evidence from both grid-connected PV adoption in Western contexts \citep{Graziano2019, Graziano2015, bollinger2012} and agricultural technology adoption in off-grid settings \citep{Conley2010}. In both domains, peer effects are temporally bounded, exerting their strongest influence shortly after prior adoption events and weakening as temporal distance increases. This pattern can be explained through social learning mechanisms, which enable potential adopters to acquire information and practical experience from earlier adopters \citep{xiong2016peer, Rogers2014}. These mechanisms are likely to be most effective shortly after an adoption event, when a newly installed system attracts attention and can facilitate direct knowledge exchange. The event itself may temporarily increase both adopters' motivation to share recent experiences and potential adopters' interest in seeking information. Over time, however, the event becomes routinized within the landscape, its novelty fades, and the opportunities for direct experience sharing may diminish, reducing the marginal influence of installation on subsequent adoption decisions \citep{BalaGoyal1998, Young2009}. 

Our second observation, that SSC strengthens over calendar time, contrasts with evidence from Western grid-connected sites, where peer effects tend to peak early in the adoption process and weaken as diffusion progresses and matures \citep{bollinger2012, Graziano2015, Graziano2019, Rogers2014}. This divergence may stem from the slower rate of PV diffusion in off-grid communities. In these settings, adoption often unfolds incrementally rather than through a single installation event \cite{GOGLA2023}. Households gradually expand their systems by adding panels over time as confidence in the technology grows and financial resources permit \citep{lemaire2018solar, mahieu2025}. Consequently, maturation of the adoption process may take longer than in Western economies, where PV adoption typically occurs through a one-time purchase, which is often supported by grid-related financial incentives, such as feed-in tariffs \citep{arnold2022prices}. 
Since exposure to peer behavior becomes effective only once households reach a stage of practical readiness, in more gradual off-grid PV diffusion, contagion may intensify later in the adoption trajectory rather than at its outset. This perspective is consistent with contagion models in which individuals transition between susceptible and adopter states as exposure accumulates through social contacts \citep{Herrera2015, Young2009}. Hence, readiness to adopt is not static but evolves as social interactions and external influences reshape the opportunity structure for adoption, a mechanism that may help account for the observed intensification in contagion as the adoption process unfolds.

Our findings suggest that spatial scale and temporal dynamics of SSC should inform the design of electrification policy in off-grid contexts. Since PV diffusion in these settings operates at the community scale, the community itself should be treated as the primary unit of intervention rather than the individual household. Programs that identify and engage influential adopters within the community, while strengthening the circulation of practical knowledge and user experience, may therefore amplify adoption. The temporal patterns identified here further suggest that policy engagement should extend beyond short-term deployment initiatives. Because adoption in off-grid contexts often unfolds gradually and becomes more feasible as households accumulate resources and familiarity with the technology, sustained policy support may be required to accompany communities over the longer adoption trajectory.

\subsection{SSC patterns and adoption outcomes}
The second objective was to identify distinct SSC patterns, quantified by SSC intensity and range, and examine their association with adoption outcomes. This research approach shifts the focus from average effects to structurally distinct configurations, thereby revealing heterogeneity that aggregate estimates would otherwise obscure.

Our findings indicate that SSC heterogeneity is not merely descriptive but substantively linked to the adoption process in off-grid communities. In particular, they suggest that SSC intensity and range provide distinct insights into adoption dynamics: the former is more closely associated with cumulative adoption performance, whereas the latter becomes especially relevant for understanding how adoption expands within households over time.
A central insight is that the greatest differentiation in cumulative adoption occurs along the SSC intensity dimension. Communities with stronger SSC consistently achieved higher overall adoption outcomes, regardless of whether contagion operates over a longer or shorter spatial reach. This suggests that the magnitude of localized reinforcement may matter more for long-run adoption performance than the breadth of clustering alone. This distinction is theoretically important. Much of the literature on peer effects in PV adoption has emphasized effective distance or decay, with proximity as the principal descriptor of contagion \citep{barton2021decay,  BaltaOzkan2021}. Our findings suggest that this approach is incomplete. Two communities may exhibit comparable spatial reach of exposure yet differ substantially in adoption outcomes if the local reinforcing effect of that exposure differs in strength. SSC intensity and SSC range should therefore be treated as analytically distinct rather than interchangeable dimensions of diffusion.

The role of SSC range becomes more evident when adoption is examined through an incremental expansion of PV systems within households over time. Even among communities with similar overall adoption outcomes, a broader range of contagion is associated with greater household adoption expansion. Thus, the spatial breadth of contagion relates not only to whether adoption occurs, but also to the extent of its expansion over time. This distinction is especially meaningful in off-grid, resource-constrained settings, where PV systems are rarely installed at full capacity in a single step, but instead evolve incrementally as households progressively increase system capacity and broaden the range of electricity uses over time \citep{lemaire2018solar, mahieu2025, GOGLA2023}. In these settings, expansion over time is not a secondary outcome, but part of the core adoption trajectory itself.

The observed link between SSC patterns and the expansion of household adoption may suggest a reinforcing cross-scale dynamic. In communities where SSC is both intense and spatially extensive, households are not only more likely to undertake initial adoption but also more likely to scale up their systems over time. Moreover, these incremental expansions may themselves strengthen local signals of viability and reliability, thereby sustaining visibility and reinforcing SSC. Rather than reflecting a unidirectional effect, the coupling between SSC intensity, range, and within-household expansion may point to a mutually reinforcing process in which household-level scaling and community-level diffusion evolve in tandem. Although the evidence remains correlational, the consistency of this alignment across patterns suggests that these processes are systematically linked rather than coincidental.

This configuration is consistent with theories of endogenously reinforced diffusion \citep{Bass1969, Young2009, Rogers2014}, in which adoption is amplified by socially and spatially internal feedback dynamics rather than driven solely by exogenous exposure. As households progressively expand their systems, visible signals of viability accumulate, strengthening subsequent adoption responses. Such dynamics are closely linked with mechanisms of complex contagion \cite{centola2007} and conceptually consistent with notions of increasing returns \citep{arthur1994} in the diffusion literature: repeated and validated exposures amplify adoption responses beyond what isolated signals would generate. These feedback processes also imply non-linear diffusion trajectories, with communities potentially diverging over time depending on the strength of reinforcing dynamics \citep{granovetter1978}.

Our findings also explain why peer-effect estimates may vary so widely across different contexts. Aggregate or radius-based measures alone may obscure the fact that communities differ not only in how far contagion extends, but in whether that exposure is sufficiently reinforcing to generate continued adoption gains. The typology developed here, therefore, provides a more differentiated framework for understanding socio-spatial diffusion in resource-constrained settings.

\subsection{Transitions in SSC patterns}
The third objective extends the analysis from static SSC patterns to their evolution over the adoption process. Overall, SSC patterns display substantial persistence across the adoption process. Most communities in the studied site maintain their relative contagion configuration as the regional trend strengthens, with their absolute levels of SSC intensity and range continuing to rise. This indicates that contagion remains an active organizing mechanism long after the onset of diffusion. In light of the broad literature, which often documents that contagion mechanisms are largely confined to early phases of the adoption process and weaken as diffusion unfolds \citep{Sokolowski2023, barnes2022passive, Graziano2015}, this persistence may imply a slower, more protracted diffusion trajectory.

Yet, a meaningful subset of communities undergoes transitions in SSC patterns, either accelerating or decelerating contagion over time, with these shifts carrying different meanings at different adoption phases. In early phases, transitions in SSC appear to track the direction of adoption dynamics, indicating whether diffusion is gaining momentum or struggling to sustain adoption growth. Communities with stronger growth in adoption intensity also exhibit acceleration in their SSC, whereas communities with weaker adoption gains show a corresponding deceleration. This may suggest that in this phase, prior installations play a central role in shaping subsequent diffusion. Adoption and contagion growth, therefore, appear to unfold together.

This observation is theoretically consistent with the environmental settings characterizing early stages of PV penetration, when installations are still sparse, and their high visibility may generate a strong spatial signal that extends beyond the immediate vicinity \citep{rode2020spot, Graziano2015, Min2025}. This effect can be pronounced in off-grid rural settings, where the built environment is low-rise and spatially compact \citep{opiyo2019impacts}. A similar visual salience effect of early PV installations was also documented in rural grid-connected contexts in Western countries \citep{Min2025}.

Over the course of diffusion, the meaning of SSC range transitions changes, and strong gains in adoption intensities no longer coincide with further contagion expansion. Instead, communities experiencing high adoption growth tend to shift toward more spatially contracted contagion patterns. This pattern transition suggests a reorganization of PV installations' spatial structure toward a more consolidated distribution, which may be explained by the gradual normalization of the technology within the community. As the feasibility and benefits of the technology gain broader recognition, the linkage between exposure, observational learning, and imitation could weaken, and adoption decisions may become less dependent on visual exposure. Under such conditions, new installations may emerge more broadly across the community space, reducing long-range clustering. The observed contraction in contagion range within high-adoption communities aligns with the broader literature, which documents a decay in the spatial influence of prior PV installations as the adoption process unfolds \citep{barnes2022passive, Graziano2015, rode2020spot}. 

Yet, late-phase range contraction found in strongly adopting communities should not be interpreted as an erosion of contagion. Alongside this contraction, SSC intensity continues to increase and remains positively aligned with adoption growth. As contagion intensity increases while its spatial range narrows, peer influence is likely to be more concentrated and intense within the immediate local environment. Under such conditions, social learning is likely to occur through more active forms of interpersonal interaction, such as word-of-mouth communication among nearby households, which can serve as an important channel for information circulation.

This interpretation engages with several studies that distinguish between passive peer effects, operating through visual exposure, and active peer effects, operating through interpersonal encounters and direct exchange of information \citep{morrissey2024takes, barton2021decay, rai2013effective}. Contagion may therefore operate through different mechanisms and remain influential even as its spatial manifestation becomes more localized. The perspective is consistent with some evidence suggesting that the relative importance of these mechanisms may vary across adoption stages \citep{scheller2021active}, with passive peer effects appearing more pronounced in early stages of diffusion than in later ones \citep{barnes2022passive}. This pattern is also aligned with theoretical perspectives on complex contagion \citep{centola2007, Young2009}, which emphasize that adoption often requires ongoing reinforcement from multiple sources rather than exposure alone. As visual exposures become less informative over time, reinforcement through local networks may become increasingly important.

Our findings may also indicate substantial disparities in diffusion trajectories across the study area, a pattern likewise observed in other off-grid regions  \citep{Aarakitl2021, saha2025}. While certain localized pockets display contagion dynamics consistent with more advanced adoption stages, the continued strengthening of SSC across a large part of the study area suggests that adoption remains comparatively less mature. A possible explanation for these disparities lies in slower and more incremental adoption processes characteristic of many off-grid contexts \citep{mahieu2025,zaman2021impact}. As PV uptake in such settings often unfolds over extended periods, it is plausible that only a limited subset of communities reaches relatively advanced adoption stages within the decade following the onset of diffusion examined here, while many others may still remain in a phase of ongoing development.

SSC intensity and SSC range, therefore, capture distinct yet interrelated and evolving dimensions of contagion dynamics that both shape the adoption process and manifest differently along its course. Their joint evolution indicates that contagion mechanisms do not necessarily weaken but rather reorganize. Accordingly, the two should be considered analytically separate and complementary measures of contagion dynamics.

Electrification efforts and strategies to promote solar energy adoption in off-grid contexts may benefit from accounting for the spatio-temporal dynamics of SSC. Attending to the co-evolution of its distinct dimensions and their associated spatial scales throughout the diffusion process may support more targeted, context-sensitive interventions, for example, through strategic seeding of installations to leverage contagion dynamics to promote adoption.

\section{Conclusion}
This study provides a data-driven analysis of SSC dynamics in off-grid residential PV adoption. By leveraging deep-learning segmentation of high-resolution aerial imagery, we reconstruct household-level installation events and quantify spatio-temporal structures using point-process inference. Our findings suggest that PV adoption in off-grid communities is shaped by contagion dynamics operating at multiple spatial and temporal scales. Across 507 communities, SSC is nearly ubiquitous, with its spatial reach often extending beyond immediate neighbors to encompass much of the community. SSC is temporally concentrated, peaking within one to two years after nearby exposure and attenuating at longer lags, while also strengthening as the adoption process progresses. Both the intensity and the spatial range of SSC are positively associated with community-level cumulative adoption outcomes and with incremental growth in household-level adoption over time.

Beyond establishing these regularities, our analysis demonstrates that SSC dynamics evolve as the adoption process advances. In the early phases, growth in both SSC intensity and its spatial reach is consistently associated with faster adoption growth. In later phases, adoption growth increasingly coincides with a deceleration in SSC reach alongside further strengthening of SSC intensity. These dynamics suggest a transition toward a more community-scale–independent adoption pattern, characterized by increasing spatial consolidation, while SSC continues to operate at the immediate-neighbor scale.

The results underscore the value of integrating data-driven behavioral records with spatio-temporal inference to refine both diffusion research and context-sensitive policy design in underserved off-grid regions.

\subsection{Limitations and future research}
This study has several limitations, some of which also highlight avenues for future research. First, SSC is inferred from spatio-temporal clustering patterns derived from point pattern analysis. While Ripley-based methods capture spatial dependence consistent with SSC, they do not establish causal peer effects, as observed patterns may also reflect shared environmental or contextual factors. The findings should therefore be interpreted as indicative rather than direct causal influence. Second, while our analysis documents systematic heterogeneity in SSC extent and dynamics, it does not identify the underlying factors that generate these differences across communities. Future work may integrate socio-economic, institutional, and infrastructural determinants to explain why some communities exhibit stronger, more persistent, or more rapidly evolving SSC than others. Third, although the spatio-temporal point-process framework captures observed adoption behavior with high spatial and temporal resolution, it cannot disentangle peer influence from unobserved correlated shocks or shared constraints that may also shape adoption decisions. Combining observational data with qualitative field evidence could establish causal and systemic interpretation. Fourth, our analysis focuses on a single off-grid context, which provides a theoretically grounded case but may not fully capture the diversity of off-grid settings. Extending the approach to additional regions with differing social structures, settlement morphologies, and policy environments would help assess the robustness and transferability of the identified SSC mechanisms. 

\section*{Acknowledgment}
The authors thank the annotators who participated in the project. This work was supported by Israel Science Foundation (ISF) under Grant 299/23.

\bibliographystyle{model1-num-names}
\bibliography{refs}

\FloatBarrier
\vfill
\break
\appendix
\section*{}
\setcounter{table}{0}
\renewcommand{\thetable}{A\arabic{table}}
\renewcommand{\tablename}{Appendix}

\begin{table}[H]
\centering
\footnotesize
\caption{SAM 3 performance metrics by year}
\label{model}
\renewcommand{\arraystretch}{1.05}
\begin{tabular*}{0.9\columnwidth}{@{\extracolsep{\fill}}
>{\hspace{6pt}\centering\arraybackslash}p{1cm}
>{\centering\arraybackslash}p{1cm}
>{\centering\arraybackslash}p{1cm}
>{\centering\arraybackslash}p{1cm}
>{\centering\arraybackslash\hspace{-1pt}}p{1cm}
@{}}
\toprule
\textbf{Year} & \textbf{Precision} & \textbf{Recall} & \textbf{F1} & \textbf{IoU} \\
\midrule
2012 & 0.856 & 0.869 & 0.863 & 0.759 \\
2015 & 0.869 & 0.836 & 0.852 & 0.743 \\
2016 & 0.917 & 0.901 & 0.909 & 0.833 \\
2017 & 0.931 & 0.947 & 0.939 & 0.885 \\
2020 & 0.929 & 0.945 & 0.937 & 0.881 \\
2021 & 0.947 & 0.952 & 0.950 & 0.904 \\
2022 & 0.976 & 0.946 & 0.961 & 0.924 \\
\bottomrule
\end{tabular*}
\end{table}

\begin{table}[H]
\centering
\footnotesize
\caption{Descriptive statistics of detected PVs by year}
\label{panels}
\renewcommand{\arraystretch}{1.01}
\begin{tabular*}{0.9\columnwidth}{@{\extracolsep{\fill}}
>{\hspace{6pt}\centering\arraybackslash}p{0.85cm}
>{\centering\arraybackslash}p{2cm}
>{\centering\arraybackslash}p{0.8cm}
>{\centering\arraybackslash\hspace{2pt}}p{0.8cm}
@{}}
\toprule
\textbf{Year} & \textbf{Total panels} & \multicolumn{2}{c}{\textbf{Panel size \scriptsize{(m$^2$)}}} \\
 & (count) & Mean & SD \\
\midrule
2012 & 4,847  & 4.31 & 2.55 \\
2015 & 8,007  & 4.94 & 3.14 \\
2016 & 9,324  & 5.26 & 3.62 \\
2017 & 13,514 & 5.43 & 4.11 \\
2020 & 18,036 & 6.16 & 4.57 \\
2021 & 20,987 & 6.82 & 5.16 \\
2022 & 25,354 & 8.27 & 6.24 \\
\bottomrule
\end{tabular*}
\end{table}

\end{document}